\documentclass[12pt]{article}
\usepackage{graphicx}
\usepackage{subfig}

\newcommand\pubnumber{CMS experiment}
\newcommand\pubdate{\today}

\def\institute{Instituto de Ciencias y Tecnologías Espaciales de Asturias (ICTEA)\\
Universidad de Oviedo}

\def\Title#1{\begin{center} {\Large #1 } \end{center}}
\def\Author#1{\begin{center}{ \sc #1} \end{center}}
\def\Address#1{\begin{center}{ \it #1} \end{center}}

\newcommand\pubblock{\rightline{\begin{tabular}{l} \pubnumber\\
         \pubdate  \end{tabular}}}
\newenvironment{Abstract}{\begin{quotation}  }{\end{quotation}}
\newenvironment{Presented}{\begin{quotation} \begin{center} 
             PRESENTED AT\end{center}\bigskip 
      \begin{center}\begin{large}}{\end{large}\end{center} \end{quotation}}

\def\beq{\begin{equation}}
\def\eeq#1{\label{#1}\end{equation}}
\def\eeqn{\end{equation}}

\def\beqa{\begin{eqnarray}}
\def\eeqa#1{\label{#1}\end{eqnarray}}
\def\eeqan{\end{eqnarray}}

\let\bar=\overbar

\def\Dslash{\not{\hbox{\kern-4pt $D$}}}
\def\dslash{\not{\hbox{\kern-2pt $\del$}}}

\def\msb{{\bar{\ssstyle M \kern -1pt S}}}

\newcommand{\ttbar}{$\mathrm{t}\bar{\mathrm{t}}$ }
\newcommand{\emu}{$\mathrm{e}^\pm \mu^\mp$ }
\newcommand{\pt}{$p_{\mathrm{T}}$ }

\begin{document}
\begin{titlepage}
\pubblock

\vfill
\Title{Inclusive and differential cross-sections measurements in the single top tW e-mu channel with CMS}
\vfill
\Author{ Alejandro Soto Rodríguez for the CMS Collaboration}
\Address{\institute}
\vfill
\begin{Abstract}
Inclusive and normalised differential cross sections measurements are presented for the production of single top quarks in association with a W boson, in proton-proton collisions at a centre-of-mass energy of 13 TeV. Events containing one muon and one electron in the final state are analysed. For the inclusive measurement, a multivariate discriminant, exploiting the kinematic properties of the events, is used to separate the signal from the dominant t$\bar{\mathrm{t}}$ background. For the differential measurements, a fiducial region is defined according to the detector acceptance, and the requirement of exactly one b-tagged jet. The resulting distributions are unfolded to particle-level and compared with predictions at next-to-leading order in perturbative QCD. Within current uncertainties, all predictions agree with the data.
\end{Abstract}
\vfill
\begin{Presented}
$14^\mathrm{th}$ International Workshop on Top Quark Physics\\
(videoconference), 13--17 September, 2021
\end{Presented}
\vfill
\end{titlepage}
\def\thefootnote{\fnsymbol{footnote}}
\setcounter{footnote}{0}

\section{Introduction}
Single top quarks were observed for the first time by the D0~\cite{D0} and CDF~\cite{CDF} Collaborations at the Fermilab Tevatron collider. There are three main production modes in proton-proton (pp) collisions: the production and decay of a virtual W boson (\textit{s}  channel), the exchange of a virtual W boson (\textit{t} channel), and the associated production of a top quark and a W boson (tW channel).

The tW process interferes with top quark pair production ($\mathrm{t}\bar{\mathrm{t}}$) at NLO in QCD. In this analysis, two schemes are defined to treat this interference: diagram removal (DR), in which all doubly resonant diagrams are removed from the matrix element (ME) calculation, and diagram subtraction (DS), in which a gauge invariant term is introduced in the ME calculation that locally cancels the doubly resonant diagrams. The tW production cross section is computed at approximate next-to-next-to-leading order (NNLO). The corresponding theoretical prediction for the tW cross section in pp collisions at $\sqrt{s} = 13$ TeV, assuming a top quark mass ($m_{\mathrm{t}}$) of 172.5 GeV~\cite{TheoXsec}, is $\sigma^{\mathrm{SM}}_{\mathrm{tW}} = 71.7 \pm 1.8 (\mathrm{scale}) \pm 3.4 (\mathrm{PDF}) \ \mathrm{pb.}$

This document reports the first measurement from the CMS Collaboration of the inclusive \cite{tWIncl} and differential \cite{tWDiff} cross sections of the tW process in pp collisions at $\sqrt{s} = 13$ TeV. The measurement is performed using final states with one electron and one muon of opposite charge. The analysed data was recorded by the CMS detector~\cite{CMSdet} during 2016 and corresponds to $35.9 \mathrm{ fb}^{-1}$ of integrated luminosity.

\section{Event selection}
The analysis uses the \emu channel of the tW process, which corresponds to a final state with one electron and one muon of opposite charge, the two corresponding neutrinos and a jet resulting from the fragmentation of the bottom quark. The main background contribution arises from the \ttbar process, which can give very similar final states with only one additional b quark. To select signal events and reduce the background contributions an event selection is performed. 

Events are required to pass either a dilepton or single-lepton trigger. Leptons (electrons or muons) are required to be well isolated and to have \pt$\ > 20$ GeV and $|\eta|<2.4$. In events with more than two leptons passing the selection, the two with the largest \pt are selected for further study. Jets are reconstructed using the anti-$\kappa_\mathrm{T}$ clustering algorithm with a distance parameter of 0.4. Jets are required to have \pt $\ > 30 $ GeV, and $|\eta|<2.4$. To avoid double counting issues, jets within a cone of $\Delta R = 0.4$ with respect to the selected leptons are not considered. An additional category of jets called ``loose jets'' is defined to be jets passing the above selection but with \pt between 20 and 30 GeV.

Some sources of background, such as Drell-Yan (DY), do not contain jets originated from a b quark. The identification of these jets results in a significant reduction in background. Jets are identified as b jets using the algorithm CSVv2 \cite{CSVv2}.

\begin{figure}[htpb!]
    \centering
    \subfloat{\includegraphics[width = 0.45\textwidth]{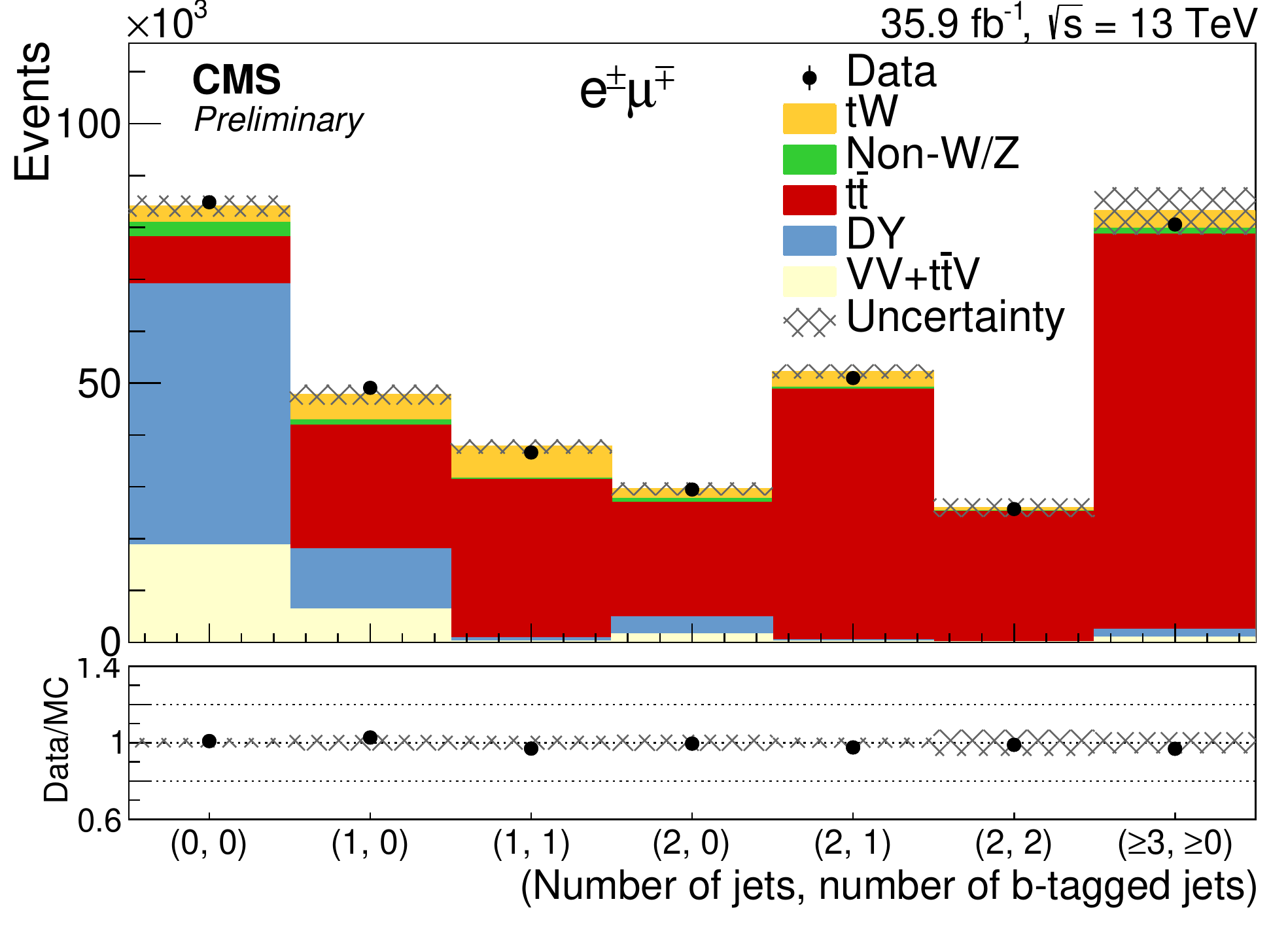}}
    \hspace{1cm}
    \subfloat{\includegraphics[width = 0.375\textwidth]{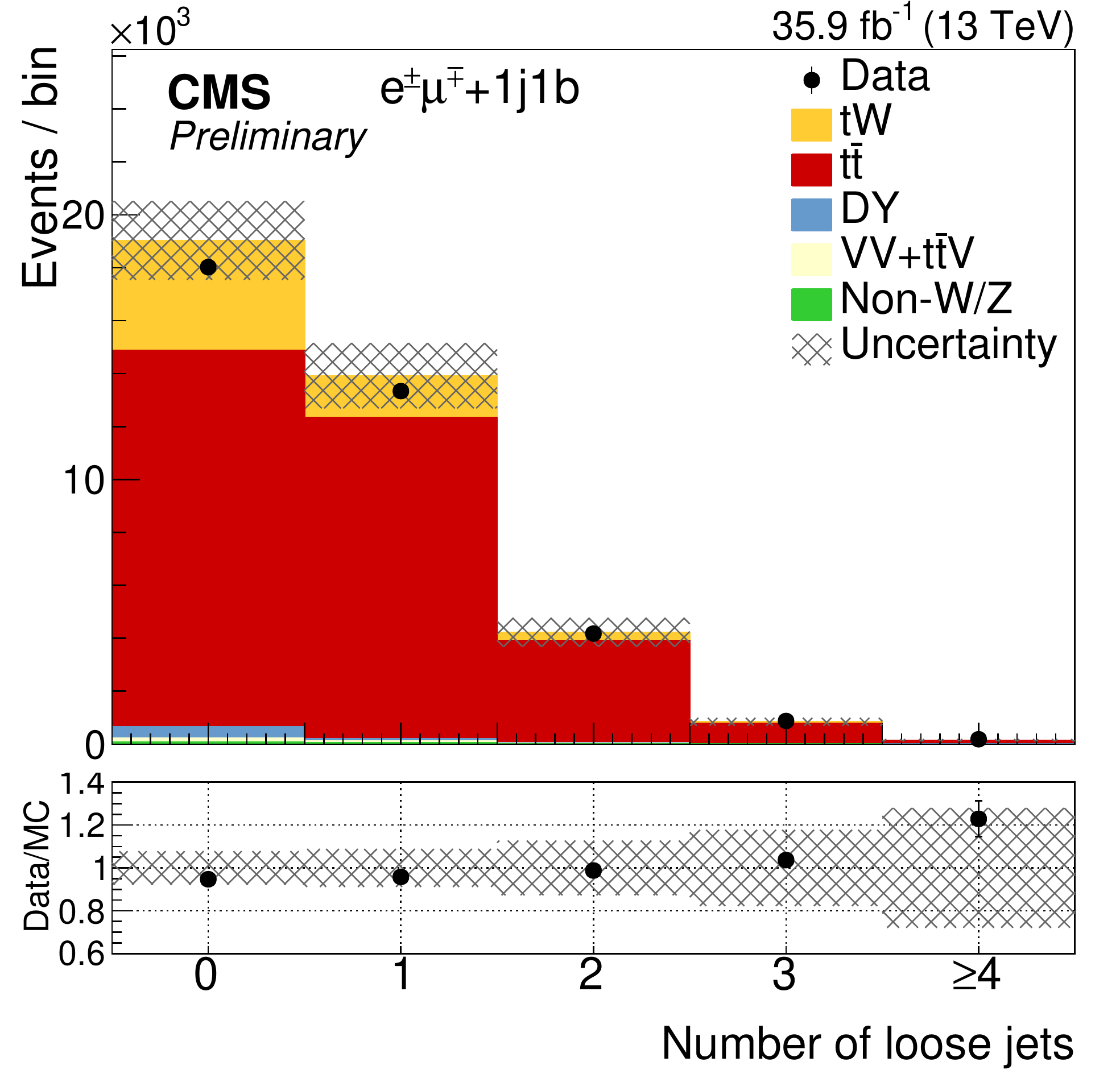}}
    \caption{Yields observed in data, compared with those expected from simulation, as a function of the number of jets and number of b-tagged jets for events passing the baseline dilepton selection (left) and as a function of the number of loose jets passing also the 1j1b selection (right). The error band includes the statistical and all systematic uncertainties, except those from background normalisation. The bottom of each panel shows the ratios of data to the sum of the expected yields \cite{tWIncl,tWDiff}.}
    \label{fig:njetsnbjetsNloose}
\end{figure}

Events are classified as belonging to the \emu final state if the two leptons with larger \pt (leading leptons) passing the above selection criteria are an electron and a muon of opposite charge. We also require that the leading lepton has \pt $\ > 25$ GeV. To reduce the contamination from low mass resonances, the invariant mass of the dilepton pair must be greater than 20 GeV. Different regions for the inclusive and differential measurements are defined based on the number of jets and b-tagged jets. Figure~\ref{fig:njetsnbjetsNloose} shows this distribution and the data/Monte Carlo (MC) comparison (left). For the inclusive measurement the regions with one b-tagged jet (1j1b), two jets and one of them b-tagged (2j1b) and two b-tagged jets (2j2b) are used, while for the differential measurements, only the 1j1b region is used with an additional selection in the number of loose jets, Fig.~\ref{fig:njetsnbjetsNloose} (right), rejecting events with one or more loose jets.

\section{Inclusive measurement}
Following the event selection the data sample in the 1j1b region consists primarily of \ttbar events with a significant number of tW events. Given that there is no single observable that discriminates between \ttbar and tW, two boosted decision trees (BDT) are trained, one in the 1j1b region and the other in the 2j1b region, to discriminate between them. The BDT implementation is provided by the TMVA~\cite{tmva} package. The 2j2b region is used as a \ttbar control region to constrain this main source of background. 

\begin{figure}[htpb!]
\centering
\subfloat{\includegraphics[width = 0.39\textwidth]{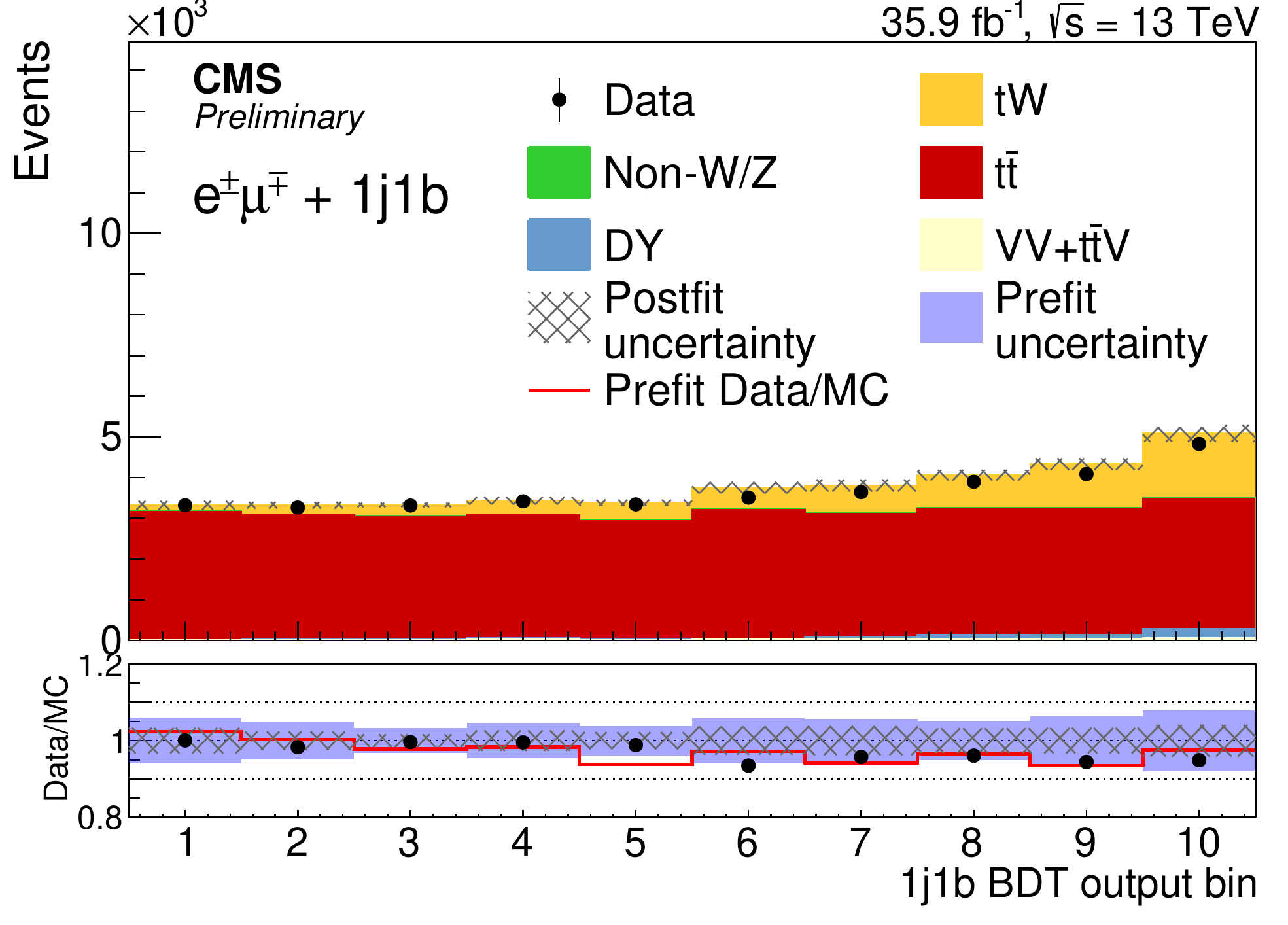}} 
\subfloat{\includegraphics[width = 0.39\textwidth]{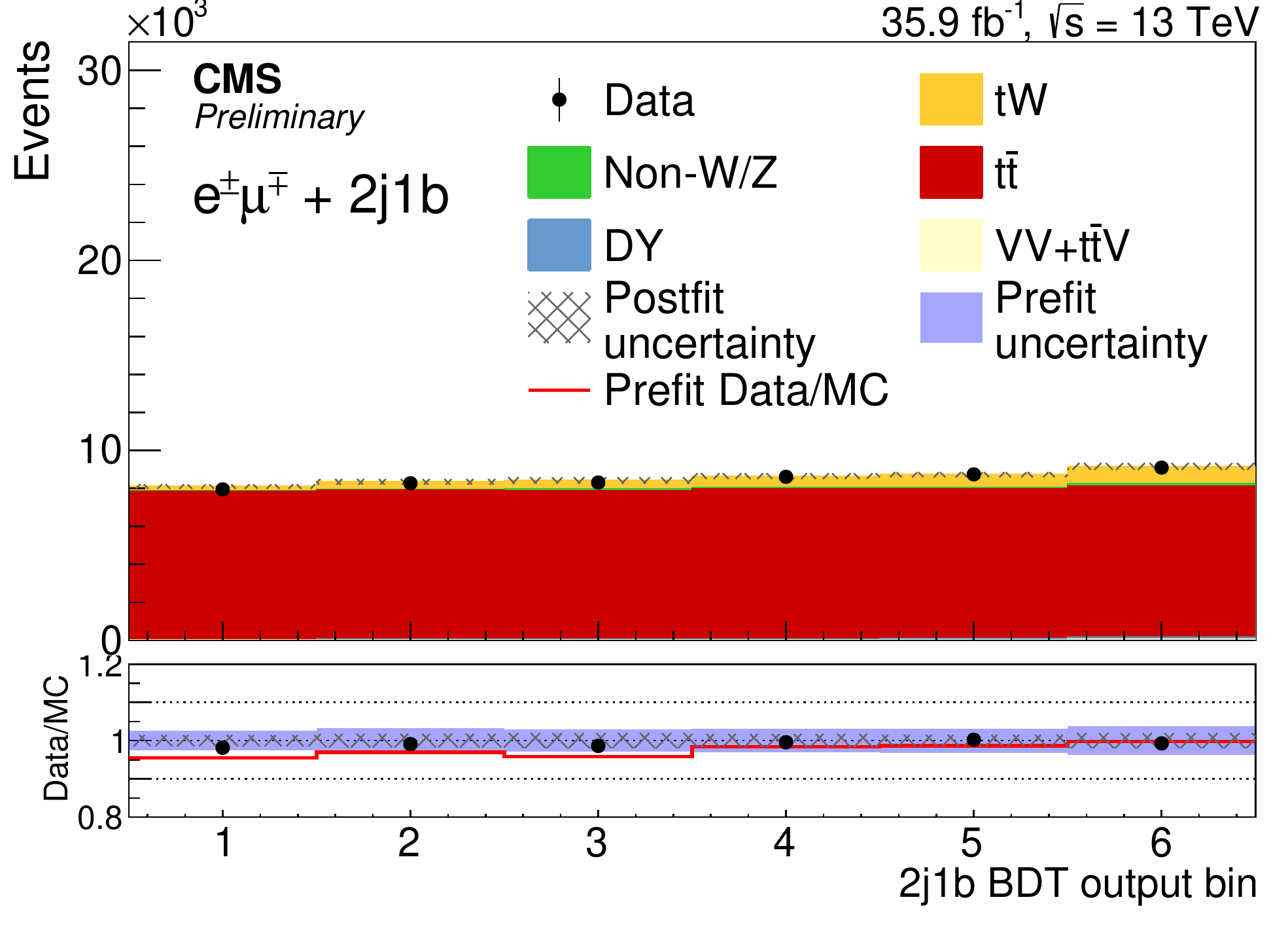}}\\
\subfloat{\includegraphics[width = 0.39\textwidth]{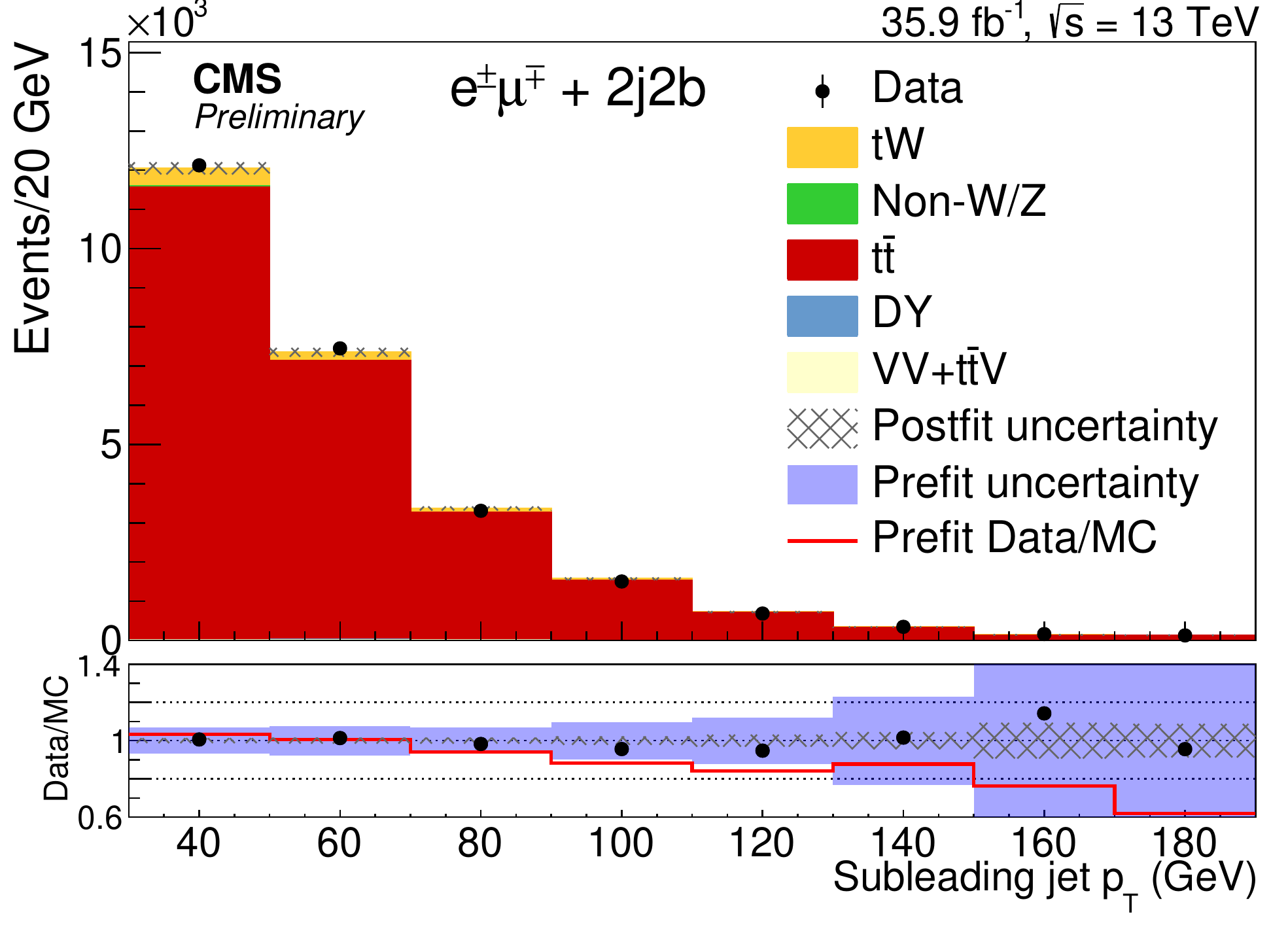}}
\caption{Comparison of the BDT output in the 1j1b (upper left) and 2j1b (upper right) regions and the \pt of the subleading jet in the 2j2b region (lower) distributions after the fit is performed for the observed data and simulated events \cite{tWIncl}.}
\label{fig:plot2}
\end{figure}

The signal is extracted by performing a simultaneous maximum likelihood (ML) fit to the distributions of the BDT output in the 1j1b and 2j1b regions and the subleading jet \pt in the 2j2b region. The binning of the BDT outputs is chosen such that each bin contains approximately the same amount of \ttbar events. This ensures that enough background events populate all the bins, helping to constrain the systematic uncertainties. Figure~\ref{fig:plot2} shows the postfit distributions of the three variables that enter in the fit.

The measured inclusive cross section of the tW process that results in the best fit to the data is $\sigma^{\mathrm{exp}}_{\mathrm{tW}} = 63.1 \pm 1.8 (\mathrm{stat}) \pm 6.4 (\mathrm{syst}) \pm 2.1 (\mathrm{lumi}) \ \mathrm{pb,}$ consistent with the SM expectations.

\section{Differential measurement}
For the differential measurement, the signal is extracted by subtracting background to data. Unfolding techniques are used to take into account migration effects when extrapolating to the fiducial region at particle level. The fiducial region is defined by the same selection requirements employed in the event selection but applied on particle-level objects. The differential cross sections are measured as function of the leading lepton \pt, jet \pt, $\Delta \phi (\mathrm{e}^\pm,\mu^\mp)$, $p_{\mathrm{z}}(\mathrm{e}^\pm,\mu^\mp, j)$, \textit{m}$(\mathrm{e}^\pm,\mu^\mp, j)$ and $\mathrm{\textit{m}}_{\mathrm{T}}(\mathrm{e}^\pm,\mu^\mp, j, p_\mathrm{T}^\mathrm{miss})$ (the definition of these variables can be found in \cite{tWDiff}) and normalised to the fiducial cross section. 

\begin{figure}[htpb!]
\centering
\subfloat{\includegraphics[width = 0.41\textwidth]{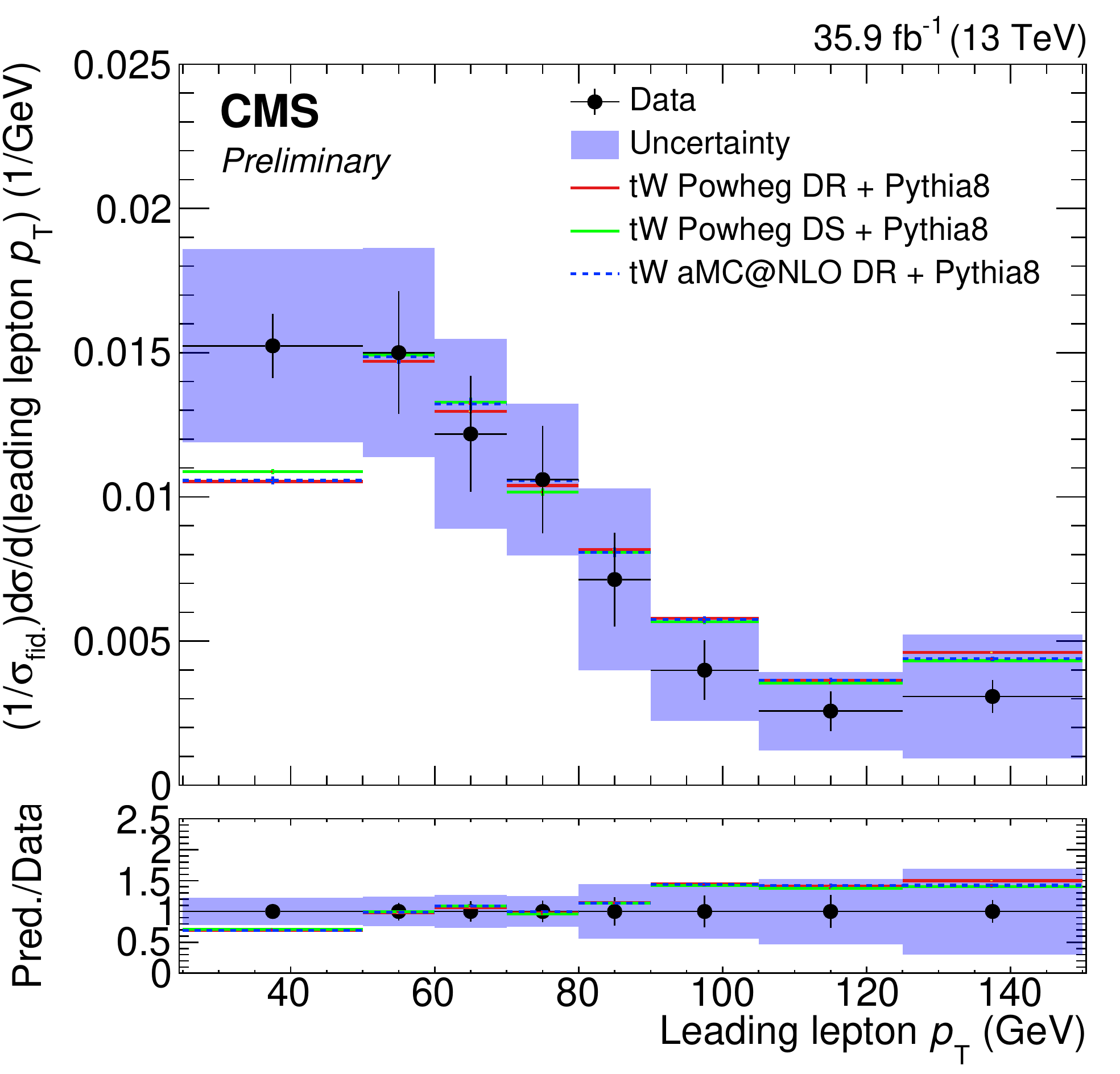}} 
\subfloat{\includegraphics[width = 0.41\textwidth]{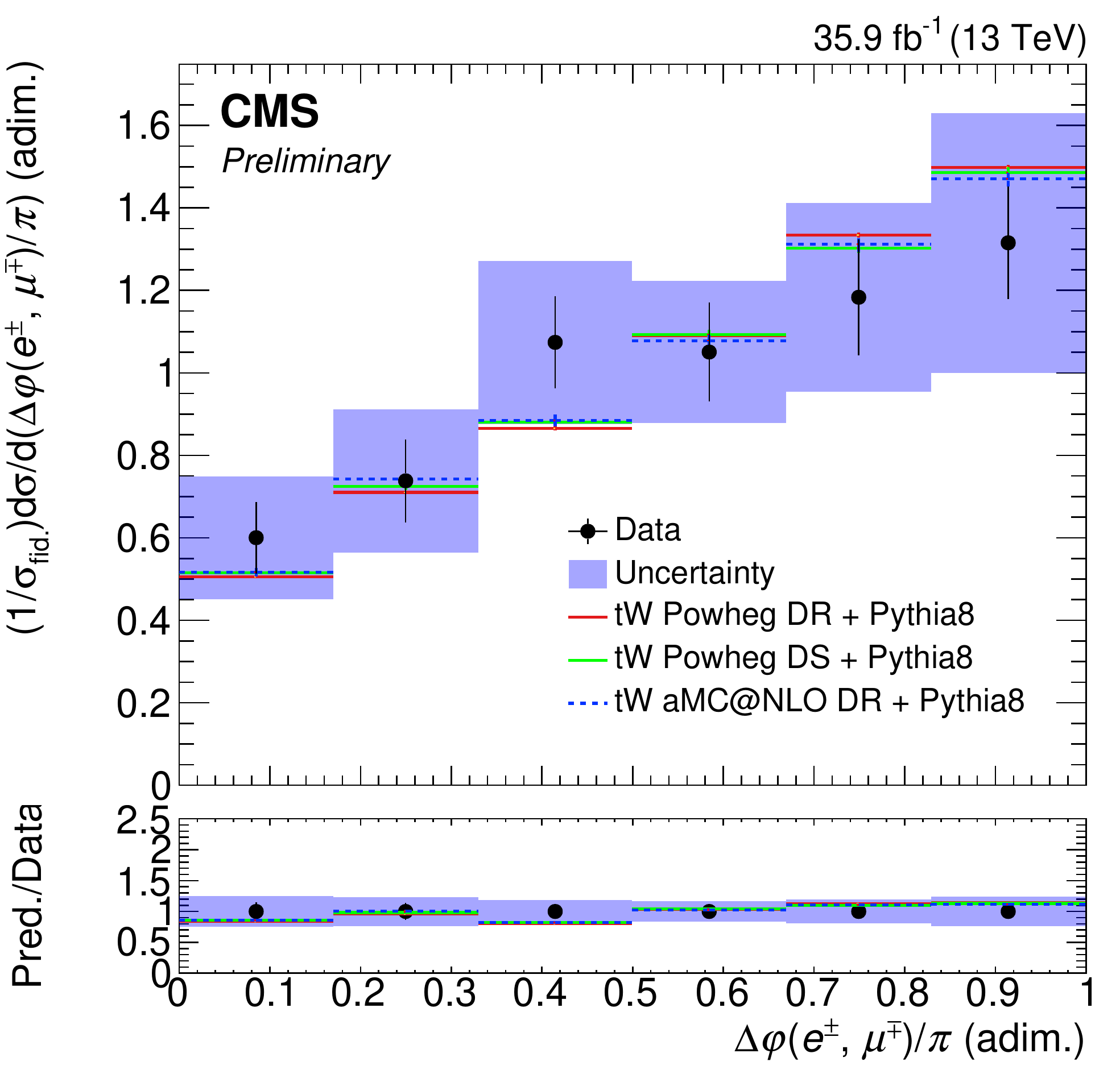}}\\
\caption{Normalised differential tW production cross section as a function of the \pt of the leading lepton and $\Delta \phi (\mathrm{e}^\pm,\mu^\mp)$ (right). The solid band represents the total uncertainty. Predictions from \texttt{POWHEG} and \texttt{MADGRAPH5\_aMC@NLO} are also shown. In the bottom panel, the ratio between predictions and data is shown \cite{tWDiff}.}
\label{fig:plot4}
\end{figure}

Figure~\ref{fig:plot4} shows the normalised differential cross sections as a function of the leading lepton \pt and $\Delta \phi (\mathrm{e}^\pm,\mu^\mp)$. Good agreement, within the uncertainties, with the predictions from \texttt{POWHEG} DR, \texttt{POWHEG} DS and \texttt{MADGRAPH5\_aMC@NLO} DR is observed.

\section{Summary}
A measurement of the inclusive and differential cross sections of the tW process at 13 TeV with the CMS detector is presented. The analysis employs the \emu channel and classifies the events in terms of the number of jets and b-tagged jets to perform the measurement. The inclusive cross section is obtained using a maximum likelihood fit to the distribution of the boosted decision tree discriminants in two categories, and the subleading jet \pt in a third category. The measured cross section is found to be $63.1 \pm 1.8 (\mathrm{stat}) \pm 6.4 (\mathrm{syst}) \pm 2.1 (\mathrm{lumi}) \ \mathrm{pb,}$ achieving a relative uncertainty of 11\% in agreement with the SM. Finally, the measured differential cross sections are, in general, consistent with the expectations from the models used to simulate the tW signal.

\bibliography{eprint}{}
\bibliographystyle{unsrt}
 
\end{document}